\documentclass[aps,prc,twocolumn]{revtex4}
\usepackage{graphicx}
\usepackage{amsmath,bbm}
\usepackage{amssymb,bm}
\usepackage{array}
\newcommand\nd{\noindent}
\begin{document}
\title{Bottomonium suppression at $\sqrt{s_{NN}}=2.76$ TeV using model based on color screening and gluonic dissociation with collisional damping}
\author{S.~Ganesh\footnote{Corresponding author:\\Email: gans.phy@gmail.com}}
\author{M.~Mishra}
\affiliation{Department of Physics, Birla Institute of Technology and Science, Pilani - 333031, INDIA}
\begin{abstract}
    We present a model to explain the bottomonium suppression in Pb$+$Pb collisions at mid-rapidity obtained from Large Hadron Collider (LHC) energy, $\sqrt{s_{NN}}=2.76$ TeV. The model consists of two decoupled mechanisms namely, color screening during bottomonium production followed by gluon induced dissociation along with collisional damping. The quasi-particle model (QPM) is used as equation of state (EOS) for the quark-gluon plasma (QGP) medium. The feed-down from higher $\Upsilon$ states, such as $\Upsilon(1P)$, $\Upsilon(2S)$ and $\Upsilon(2P)$, dilated formation times for bottomonium states and viscous effect of the QGP medium are other ingredients included in the current formulation. We further assume that the QGP is expanding according to ($1+1$)-dimensional Bjorken's boost invariant scaling law. The net suppression (in terms of $p_T$ integrated survival probability) for bottomonium states at mid rapidity is obtained as a function of centrality and the result is then compared both quantitatively and qualitatively with the recent LHC experimental data in the mid rapidity region recently published by the CMS Collaboration. We find that the current model, based on Debye color screening plus gluonic dissociation along with collisional damping, better describes the centrality dependence of bottomonium suppression at LHC energy as compared to the color screening model alone. 

\vskip 0.5cm

{\nd \it Keywords } : Color screening, Gluonic dissociation, Collisional damping, Survival probability\\
{\nd \it PACS numbers } : 12.38.Mh, 12.38.Gc, 25.75.Nq, 24.10.Pa 

\end{abstract}

\maketitle
\section{Introduction}

  Quarkonia ($Q\bar{Q}$) suppression has been considered as a potential probe to study the formation of quark-gluon plasma (QGP) and its properties for a long time. Many experimental measurements have been carried out on charmonium suppression spanning from CERN Super Proton Synchrotron (SPS)~\cite{PKSMCA,PKSRAR} to BNL Relativistic Heavy Ion Collider (RHIC)~\cite{PKSAAD} experiments. Recently the CMS Collaboration at the Larhe Hadron Collider (LHC) reported their initial measurements for the absolute $\Upsilon(1S)$ suppression as well as the relative suppression of $\Upsilon(2S) + \Upsilon(3S)$ with respect to $\Upsilon(1S)$ and they find that the excited states $\Upsilon(nS)$ are suppressed with respect to $\Upsilon(1S)$~\cite{CMS3}. They have further presented the suppression pattern of different bottomonium states separately and shown their relative ratios and finally observed the sequential suppression of bottomonium states~\cite{CMS2}. 

Charmonium suppression data in Pb$+$Pb collisions at SPS have been explained by various models~\cite{PKSSPS1,PKSSPS2,PKSSPS3,PKSSPS4,PKSSPS5,PKSSPS6,PKSSPS7}. Similarly many researchers~\cite{yunpen,zhen,rishi} have attempted to analyze the suppression data in Au$+$Au collisions at RHIC energy. However, it is seen that charmonium states are not very clean probes since significant contribution arises through the coalescence of charm and anti-charm quark pairs (regeneration effect) at RHIC and LHC energies. In contrast, the suppression pattern of bottomonium and its excited states is regarded as a cleaner probe to study the properties of matter produced in heavy-ion collision experiments because of the heavy mass of the bottom quark and the large binding energies (about two times of that of charmonia) of various bottomonium states. Due to heavier bottom quark mass as compared to charm, the cross section for the production of $b\bar b$ pairs is found to be much smaller than for $c\bar c$ pair production. Therefore, the regeneration effect via coalescence of $b\bar b$ pair has been argued to play a negligible role even at LHC energy. 
Suppression of open charm $D$ and open bottom $B$ mesons due to collisional dissociation in QGP are other potential probes.
The first results based on perturbative QCD dynamics for open charm and beauty production have been provided in Refs.~\cite{azfar, rishi2}. The authors derived the medium-induced dissociation probability for heavy $D$ and $B$ mesons moving in hot and dense QCD matter. 
 
The concept of quarkonia suppression in QGP due to Debye color screening was first advocated by Matsui and Satz in their seminal paper~\cite{mats} published in 1986. Chu and Matsui used the concept of color screening and derived a model~\cite{Chu} to analyze the $p_T$ dependence of quarkonia suppression in QGP. They had used the Bag model equation of state (EOS) for QGP in their formulation to analyze the transverse momentum dependence of charmonia suppression arising due to the Debye color screening mechanism. A few years ago, Mishra et al.~\cite{Madhu1} presented a modified version of the Chu and Matsui model of color screening by parameterizing the pressure in the transverse plane instead of energy density as was done in the original Chu and Matsui work~\cite{Chu}. The pressure parametrization is inspired by the fact that it almost vanishes (like at the surface, $r=R_T$ of cylindrically symmetric QGP formed after collisions) at the deconfined phase transition temperature $T_c$, whereas energy density has a certain finite value at $T_c$. The pressure parametrization is also able to explain the centrality dependence~\cite{Madhu1} of charmonium suppression data reasonably well. Mishra et al. still used the Bag model EOS for QGP and introduced centrality dependence in the model through Bjorken's formula for initial energy density. The accuracy of the above color screening based model has been further improved due to the work done by Srivastava et al.~\cite{Madhu2}. They employed the more realistic quasi-particle model EOS for QGP medium and the results were compared across SPS, RHIC and LHC energies at mid-rapidity, since the Bag model EOS gives a very crude EOS of QGP by assuming QGP as a noninteracting ideal gas of quarks and gluons, inconsistent with the recent observation at RHIC energy~\cite{hirano}. Furthermore, it has been shown in~\cite{PKS1,PKS2} that various thermodynamical and transport quantities obtained from the quasi-particle model EOS compare well with the recent lattice data~\cite{plumari,bors}. This emphasizes a major difference between our current version of the color screening model~\cite{Madhu2}, the modified model used by Mishra et al.~\cite{Madhu1} and the original work of Chu and Matsui~\cite{Chu}. In spite of the potential color screening mechanism, a few more suppression mechanisms are proposed to account for the observed quarkonia suppression in QGP medium. Gluonic dissociation along with collisional damping is one of them, which is suggested to be another potential mechanism playing a paramount role in explaining the observed suppression.\\  
We first employ the above color screening model to explain the bottomonia suppression at LHC energy and find that the color screening model alone is not sufficient to describe the suppression completely. Many other research groups have tried to explain the LHC data but no one is able to explain the data very satisfactorily. Work done by Laine~\cite{Laine1,Laine2,Laine3} and collaborators has shown that the effective potential between heavy quarks may have an imaginary part too. The imaginary part is due to the collisional energy imparted by the QGP. The collisional damping occurs due to the collision of bottomonia with the light hadrons in QGP. Motivated by the above facts, we propose that more than one suppression mechanism may be playing role in the suppression process. We also assume that the additional suppression comes from gluonic dissociation along with collisional damping. We use the formulation developed by Nendzig and Wolschin~\cite{Wolschin} to model the effect of gluonic dissociation and collisional damping on bottomonium states.
 
In the present work, it is assumed that the effect of color screening on bottomonium formation can be decoupled from the gluonic dissociation and collisional damping effects. After taking into account all the above-mentioned effects, we calculate the net suppression of bottomonia in the presence of QGP medium and compare the results with the recent CMS data~\cite{CMS2} on bottomonium suppression at LHC energy. 
We find that the two mechanisms of color screening and gluonic dissociation along with collisional damping have the potential to qualitatively and quantitatively describe the shape of the centrality-dependent bottomonium suppression at LHC energy. 

This paper is organized as follows. In Sec. II, we describe the color screening model in brief. In Sec. III, we give description of the gluonic dissociation and collisional damping models. Section III also describes the two different temperature models that we explore in the current work. Section IV gives the expression for net suppression in terms of survival probability after combining color screening and gluonic dissociation along with collisional damping. Section V first describes the simulation results from a purely color screening model perspective and then gives the results and discussions in more details. Finally we conclude with the important results in the Sec. VI. 
\section{Color Screening}
The precise knowledge of quarkonia production in QCD is required in order to quantify the quarkonia suppression in QGP medium. The two frequently used mechanism of bottmonia production are the color singlet model (CSM) and the color octet model (COM). Several other QCD-based approaches such as NRQCD (nonrelativistic QCD) factorization, fragmentation factorization, and $k_t$ factorization~\cite{COMpeter,Octet} have been used to theoretically determine the direct yield of quarkonia. The inclusive quarkonia production in the CDF experiment at Tevatron is found to be an order of magnitude larger than predicted by CSM. The COM is proposed to explain the above discrepancy. However, the issue of contribution due to COM is still an open question~\cite{COMpatri}. The experimental data on quarkonia suppression are described in terms of what is known as the nuclear modification factor $R_{AA}$, defined by the yield in A$+$A relative to the yield in p$+$p collisions scaled by the number of nucleon-nucleon collisions ($N_{coll}$), which is measured as a function of the collisional centrality, transverse momentum and rapidity~\cite{CMS2}. 
 A detailed theoretical estimate of quarkonia suppression employing the NRQCD approach namely, heavy quark effective theory (HQET), has been carried out in Ref.~\cite{rishi}. However, due to significant uncertainties involved in the calculation of quarkonia yield in p$+$p collisions, suppression via the nuclear modification factor approach may be less reliable and needs further cross-check. Alternatively, many phenomenological models have been proposed to directly calculate the quarkonia suppression in QGP medium without explicitly requiring the yield in p$+$p collisions. Our present work is the outcome of such an attempt.\\
We describe here a color screening model based on the quasiparticle EOS of QGP and pressure parametrization~\cite{Madhu1} in the transverse plane presented recently by Srivastava et al.~\cite{Madhu2}. The basic framework of above the color screening model was first derived by Chu and Matsui~\cite{Chu}. 
\section*{Pressure Profile}
The pressure profile of the QGP in the transverse plane is a very crucial step of the present version of the color screening formulation, which is assumed to be described by the following equation :
\begin{eqnarray}
 p(t_i,r) = p(t_i,0)\,h(r) \nonumber \\
h(r) = \left (1- \frac{r^2}{R_T^2} \right )^\beta \theta(R_T - r),
\end{eqnarray}
where $\beta=1.0$ for hard collisions in accordance with Ref.~\cite{Madhu2}. $R_T$ denotes the radius of the cylindrically symmetric plasma formed after heavy-ion collisions and it is related to the transverse overlap area $A_T$ by $R_T = \sqrt{\frac{A_T}{\pi}}$, where $A_T$ is determined by the Glauber model~\cite{PKSBAL,PKSSSA}. The factor $p(t_i,0) = (1 + \beta)\langle p\rangle_i$, where $\langle p\rangle_i$ is the average initial pressure determined by average initial energy density $\langle\epsilon\rangle_i$. The centrality dependence is introduced through Bjorken's formula~\cite{PKSBJO, PKSSSA} for average initial energy density expressed as 
\begin{equation}
\langle \epsilon\rangle_i = \frac{1}{A_T\tau_i}\frac{dE_T}{dy},
\end{equation}
where $\frac{dE_T}{dy}$ is the transverse energy deposited per unit rapidity of the output hadrons at a given centrality. It is given by $\frac{dE_T}{dy} = 1.09\,\frac{dE_T}{d\eta}$ with $\frac{dE_T}{d\eta}$ being the transverse energy per unit pseudo rapidity at a given centrality and is taken from Ref.~\cite{cms3}. The factor $1.09$ is the value of the Jacobian, which depends on the momentum distributions of produced particles at LHC energy~\cite{cms3}. At proper time $\tau=\tau_i$, $\frac{\langle p\rangle_i}{\langle e\rangle_i} = c_s^2$~\cite{PKS2}, where $c_s$ is the velocity of sound in the QGP.

The color screening model based on the QPM EOS~\cite{Madhu2} is based on the constraint that $\eta/s$ of the QGP medium varies very slowly with temperature~\cite{PKS2}, where $\eta$ and $s$ are shear viscosity and entropy density of the QGP, respectively. 
\section*{Cooling Law} 
Srivastava et al.~\cite{Madhu2} derived the following cooling law for pressure using the QPM EOS of QGP 
\begin{equation}
	p(\tau,r) = A + \frac{B}{\tau^q} + \frac{C}{\tau} + \frac{D}{\tau^{c_s^2}},
\end{equation}
where $A = -c_1$, $B = c_2c_s^2$, $C = \frac{4\eta q}{3(c_s^2 - 1)}$, and $D = c_3$. 
The constants, $c_1$, $c_2$ and $c_3$ are determined by employing boundary conditions on energy density and pressure. We take $\epsilon=\epsilon_i=\langle\epsilon\rangle_i$ at $\tau=\tau_i$ and $\epsilon=0$ at $\tau=\tau^{'}$ to compute the value of constants $c_1$ and $c_2$, where $\tau^{'}=\tau_i\,k^{-\frac{3R}{R-1}}$ ($\approx \tau_ik^{-3} for R \gg 1$), $k=T_i/T^{'}$ and $R$ is the Reynolds number for the QGP matter. Here $T^{'}$ is the temperature which corresponds to $\epsilon=0$. It is determined by the energy density versus temperature variation curve using the QPM EOS of QGP~\cite{Madhu2}. $p=p_i=\langle p\rangle_i$ at $\tau=\tau_i$, determines the value of $c_3$. As a consequence of above boundary conditions, expressions for $c_1$, $c_2$ and $c_3$ are given by

\begin{itemize}
\item $c_1 = - c_2\tau'^{-q} - \frac{4\eta}{3c_s^2\tau'}$
\item $c_2 = \frac{\epsilon_i - \frac{4\eta}{3c_s^2}\left( \frac{1}{\tau_i} - \frac{1}{\tau'} \right ) }{\tau_i^{-q} - \tau'^{-q}}$
\item $c_3 = \left (p_i + c_1 \right) \tau_i^{c_s^2} - c_2\,c_s^2 \tau_i^{-1} - \frac{4\eta}{3}\left (\frac{q}{c_s^2 - 1} \right ) \tau_i^{\left ( c_s^2 - 1\right )}$.
\end{itemize}


\section*{Constant pressure contour and radius of the screening region}
The color screening model further assumes a region inside the QGP medium, known as the screening region, where temperature is sufficiently high, more than the dissociation temperature $T_D$, so that the heavy quarkonium (bottomonium in this case) is unlikely to form inside this region or alternatively get suppressed. If, however, the quarkonia get formed in the QGP region where the temperature is below $T_D$ then they are  no longer suppressed. The time taken by the thermalized QGP, at transverse distance $r$, to cool up to the temperature $T_D$ is called the screening time $\tau_s$. It is determined by combining the cooling law of pressure and the pressure parametrization in the transverse plane.

Since we cannot solve the cooling law of pressure for $\tau$, we use a numerical approach~\cite{Madhu2} to determine screening time, $\tau_s$. Writing the above equation at initial proper thermalization time $\tau = \tau_i$ and screening time $\tau = \tau_s$ we get the following equations: 
\begin{equation}
	p(\tau_i,r) = A + \frac{B}{\tau_i^q} + \frac{C}{\tau_i} + \frac{D}{\tau_i^{c_s^2}} = p(\tau_i,0)\,h(r)
\end{equation}
and
\begin{equation}
	p(\tau_s,r) = A + \frac{B}{\tau_s^q} + \frac{C}{\tau_s} + \frac{D}{\tau_s^{c_s^2}} = p_{QGP}.
\end{equation}
Here $p_{QGP}$ is the pressure of the QGP inside screening region required to dissociate a particular $\Upsilon$ state. It is determined by employing the QPM EOS of QGP~\cite{PKS1,PKS2} using the value of $T_D$ for that $\Upsilon$ state. Finally, we equate the screening time $\tau_s$ to the dilated formation time of quarkonia, $t_F$ = $\gamma\tau_F$ (at the boundary of screening region) to find the radius of the screening region, $r_s$, where $\gamma = \frac{E_T}{M_{\Upsilon}}$ is the Lorentz factor associated with the transverse motion of the $b\bar{b}$ pair with bottomonium mass $M_{\Upsilon}$. Here $E_T=\sqrt{M_{\Upsilon}^2+p_T^2}$. Thus $\gamma$ is determined by the $p_T$ distribution ($4.0 \le p_T\le 20$ GeV)~\cite{CMS2} of the bottomonia. $\tau_F$ is the proper time required for the separation of $b\bar{b}$ to be reduced to the binding radius of the quarkonia. Hence the pair will escape the screening region and form bottomonia if $|\vec{r} + \vec{v}\,t_F| \ge r_s$ where $\vec{r}$ is the position vector at which the bottom quark-antiquark pair is created and $r_s$ is the radius of the screening region~\cite{Chu,Madhu1,Madhu2}. The above kinematic condition takes a simplified form by assuming that $\Upsilon$ is moving with transverse momentum $p_T$ in the mid-rapidity region. Thus the above escape condition can be expressed as 
\begin{equation}
	\cos \phi \ge Y, Y = \frac{(r_s^2 - r^2)m - \tau_F^2\,p_T^2/m}{2r\,\tau_F \,p_T},
\end{equation}
where $\phi$ is the angle between the transverse momentum $\vec{p}_T$ and the position vector $\vec{r}$, and $m$ = $M_{\Upsilon}$.
\section*{Survival Probability}
We assume a radial probability distribution for the production of $b\bar{b}$ pairs in hard collisions at transverse distance $r$ given by 
\begin{equation}
	f(r) \propto \left ( 1 - \frac{r^2}{R_T^2} \right )^\alpha \theta(R_T - r).
\end{equation}
The value of $\alpha$ chosen is $0.5$~\cite{Chu,Madhu1,Madhu2}.

Thus in the color screening scenario, the survival probability for the bottomonia becomes 
\begin{equation}
	S_c(p_T,N_{part}) = \frac{\int_0^{R_T} dr r f(r) \int_{-\phi_{max}}^{\phi_{max}} d\phi} {2\pi \int_0^{R_T}\,r\,f(r)\,dr}
\end{equation}
\begin{equation}
= \frac{2(\alpha + 1)}{\pi R_T^2} \int_0^{R_T} dr r \phi_{max}(r) \left \{ 1 - \frac{r^2}{R_T^2} \right \}^\alpha, 
\end{equation}
where the maximum angle $\phi_{max}$ can be expressed as:
$$
\phi_{max}(r)=\left\{\begin{array}{rl}
\pi     & \mbox{~~if $Y\le -1$}\\
\pi-\cos^{-1}|Y|  & \mbox{~~if $0\ge Y\ge -1$}\\
\cos^{-1}|Y| & \mbox{~~if $0\le Y\le -1$}\\
 0        & \mbox{~~if $Y \ge 1$} 
\end{array}\right.
$$
The above equation is slightly modified by Mishra et al.~\cite{Madhu1} as compared to the original set of equations derived by Chu and Matsui~\cite{Chu}.

\begin{table}
\caption{Values of the input data used in our simulation~\cite{abdul,Wolschin}.}
\begin{tabular}{|l|l|l|l|l|}
\hline\hline
$\Upsilon$ properties & $\Upsilon(1S)$ & $\Upsilon(2S)$ & $\Upsilon(1P)$  & $\Upsilon(2P)$ \\
\hline
Mass (GeV) & 9.46 & 10.02 & 9.99 & 10.26 \\
\hline
$\tau_F$ (fm) & 0.76 & 1.9 & 2.6 & 3.1$^*$ \\
\hline
$T_{diss}$ (MeV) & 668 & 217 & 206 & 185$^*$\\
\hline\hline
\end{tabular}
\end{table}

\begin{table}
\caption{Values of other parameters $T_i$,  $s_i$, $\alpha$, and $\beta$~\cite{Madhu2}.}
\begin{tabular}{|l|l|l|l|}
\hline\hline
 $T_i$(GeV)   & $s_i$(GeV$^3$) & $\alpha$ & $\beta$\\
\hline
 1.0  &  16.41 & 0.5 & 1.0 \\
\hline\hline
\end{tabular}
\end{table}
\section{Gluonic dissociation and collisional damping}
\section*{Collisional damping}

Following~\cite{Wolschin}, we model the singlet potential between the two quarks inside a quarkonia as  
\begin{equation} 
\begin{split}
	V(r,m_D) = \frac{\sigma}{m_D}(1 - e^{-m_D\,r}) - \\ 
\alpha_{eff} \left ( m_D + \frac{e^{-m_D\,r}}{r} \right ) - \\
i\alpha_{eff} T \int_0^\infty \frac{2\,z\,dz}{(1+z^2)^2} \left ( 1 - \frac{\sin(m_D\,r\,z)}{m_D\,r\,z} \right ),
\end{split}
\end{equation} 
where

\begin{itemize}
\item $m_D$ is the Debye mass and is given by 
\begin{math}
	m_D = T\sqrt{4\pi\,\alpha_s^T \left (\frac{N_c}{3} + \frac{N_f}{6} \right ) }, 
\end{math}
\item $\alpha_{eff} = \frac{4\alpha}{3} = 0.63$, $N_f = 3$ = number of flavours, $\alpha_s^T = 0.1184\times 2\pi\,T$, and $\sigma= 0.192$ GeV$^2$.
\end{itemize}

The imaginary part of the potential, models the collisional damping, and the expectation value $\Gamma_{damp} = \int[\psi^\dagger \left [Im(V)\right ] \psi]$\,dr gives the dissociation constant due to collisional damping. Here, $\psi$ is the bottomonium wave function.

By solving the Schr\"{o}dinger's equation we get the radial wave functions for $1S$, $2S$ and $1P$ states as depicted in Fig. 1, the same as shown in~\cite{Wolschin}. 

\section*{Gluonic dissociation}
Again following~\cite{Wolschin} we model the gluonic dissociation cross section as:
\begin{equation}
\begin{split}
\sigma_{diss,nl}(E_g) = \frac{\pi^2\alpha_s^u E_g}{N_c^2} \sqrt{\frac{m}{E_g + E_{nl}}} \\
\left ( \frac{l|J_{nl}^{q,l-1}|^2 + (l+1)J_{nl}^{q,l+1}|^2}{2l+1} \right ), 
\end{split}
\end{equation}
where $J_{nl}^{ql'}$ can be expressed using singlet and octet wave functions as:
\begin{equation}
J_{nl}^{ql'} = \int_0^\infty dr\,r \,g^*_{nl}(r)h_{ql'}(r)
\end{equation}
and 
$\alpha_s^u = 0.59$. 
\\
The octet wave function $h_{ql}$ is the wave function obtained by solving the Schr\"{o}dinger's equation with potential, $\alpha_{eff}/(8\,r)$~\cite{Wolschin,Octet}. The Schr\"{o}dinger equation has been solved by taking a $10^4$ point logarithmically spaced finite spatial grid, and solving the resulting matrix eigenvalue equations. For the octet modeling the potential is repulsive, which implies that the quark and anti-quark can be far away from each other (at infinity). To account for this, the finite spatial grid is taken over a very large distance namely $10^2$ fm, as an approximation for infinity. The octet wave functions corresponding to large quark antiquark distance have negligible contributions to the gluonic dissociation cross section. The cross section is then averaged over a Bose-Einstein distribution of gluons at temperature $T$, 
\begin{equation}
\Gamma_{diss,nl} = \frac{g_d}{2\pi^2} \int_0^\infty \frac{dp_g\,p_g^2 \sigma_{diss,nl}(E_g)}{e^{E_g/T} - 1}.
\end{equation}

The net dissociation constant is given by 
\begin{equation}
	\Gamma_{total} = \Gamma_{damp} + \Gamma_{diss}
\end{equation}
For the temperature variation, research groups have used either multiplicity~\cite{abdul} or number of participants ($N_{part}$) variants~\cite{Temp, Wolschin}. 
We explore both types of variants in the current formulation. 
\\
For the multiplicity variant~\cite{abdul} 
\begin{equation}
T(t) = T_c \frac{\left(\frac{dN_{ch}}{d\eta}/\frac{N_{part}}{2}\right)_{bin}^{1/3}} {\left(\frac{dN_{ch}}{d\eta}/\frac{N_{part}}{2}\right)_{bin0}^{1/3}} \left ( \frac{t_{QGP}}{t} \right )^{1/3}. 
\end{equation} 
For the number of participants variant of the temperature model~\cite{Temp,Wolschin} 
\begin{equation}
T(t) = T_c \left(\frac{N_{part}(bin)}{N_{part}(bin_0)}\right)^{1/3} \left ( \frac{t_{QGP}}{t} \right )^{1/3}, 
\end{equation} 
where $t_{QGP}=$ QGP lifetime.\\
The net survival probability due to gluonic dissociation along with collisional damping is then expressed as :
\begin{equation}
	S_g= S_0\,e^{\int_{t_F}^\infty -\Gamma_{total}\,dt},
\end{equation}
where $S_g$ is the survival probability of bottomonia arising due to the effect of gluonic dissociation along with collisional damping.
\section{Combining color screening, gluonic dissociation, and collisional damping}
We now combine the above mechanisms together to write down the final expression for survival probability. 
It is to be noted that color screening affects formation of quarkonia when the temperature is higher than the dissociation temperature, while gluonic dissociation and collisional damping are predominant after quarkonia is formed i.e., when the temperature is below the dissociation temperature. This supports our approach to decouple the two mechanisms. The final survival probability due to color screening and gluonic dissociation along with collisional damping is given by combining the survival probability in Sec. II and III as follows: 
\begin{equation}
	S= S_c\,S_g. 
\end{equation}
The expressions for survival probability after incorporating feed-down corrections are given by 
\begin{eqnarray}
S_{1S} = 0.6489\,S'_{1S} + 0.1363\, S'_{1P} + 0.1733 \, S'_{2S}\\ \nonumber 
+ 0.0416 \, S'_{2P}\nonumber, 
\end{eqnarray}
\begin{eqnarray}
S_{1P} = 0.8450\, S'_{1P} + 0.1508 \, S'_{2S} + \nonumber 
0.0041 S'_{2P} \nonumber,
\end{eqnarray}
\begin{eqnarray}
S_{2S} = 0.8780 \, S'_{2S} + 0.1220 \, S'_{2P}\nonumber, 
\end{eqnarray}
where $S'_{nl}$ is the $p_T$ integrated survival probability of the $|nl\rangle$ quarkonia states before feed-down is considered, while $S_{nl}$ is the survival probability of the $|nl\rangle$ state after feed-down.   

In our calculation, the values of mass ($M_{\Upsilon}$), formation time $\tau_F$ and dissociation temperature $T_D$ of bottomonium states are depicted in Table I. We use $Tc=0.170$ GeV in accordance with the recent lattice QCD results. Other parameters, such as initial temperature ($T_i$), entropy density ($s_i$) at proper time $\tau_i$ along with $\alpha$ and $\beta$ at LHC energy in accordance with~\cite{Madhu2} are tabulated in Table II.
 The $T_D$ value of $185$ MeV for $\Upsilon(2P)$ is obtained by simulation, and the formation time $\tau_F$ for the $\Upsilon(2P)$ state is taken to be $3.1$ fm (marked with $*$ in Table I) as an estimated place holder. We assume that the error would be very low, since the feed-down contribution due to $\Upsilon(2P)$ is very small. The $p_T$ integrated survival probability $S_{\Upsilon}$ is determined over a range $4.0\le p_T \le 20$ GeV supported by the CMS experimental data~\cite{CMS2}.

\begin{figure}[h]
\includegraphics[width = 80mm,height = 80mm]{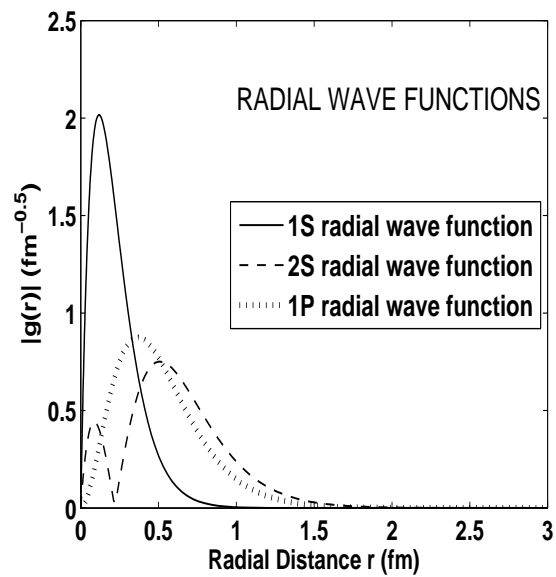}
\caption{Radial wave functions for $1S$, $2S$, and $1P$ bottomonium states at $0.2$ GeV.}
\label{fig:radial}
\end{figure}

\begin{figure}[h]
\includegraphics[width = 80mm,height = 80mm]{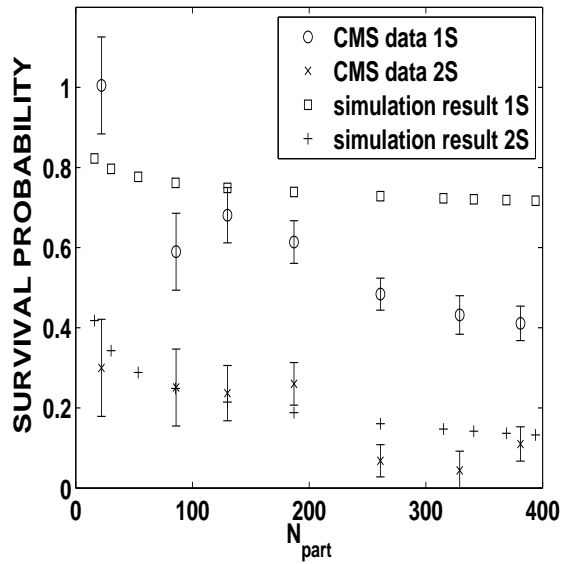}
\caption{CMS data~\cite{CMS2} compared with present simulation results. The simulation results include only the color screening mechanism.}
\label{fig:1S_2S}
\end{figure}

\begin{figure}[h]
\includegraphics[width = 80mm,height = 80mm]{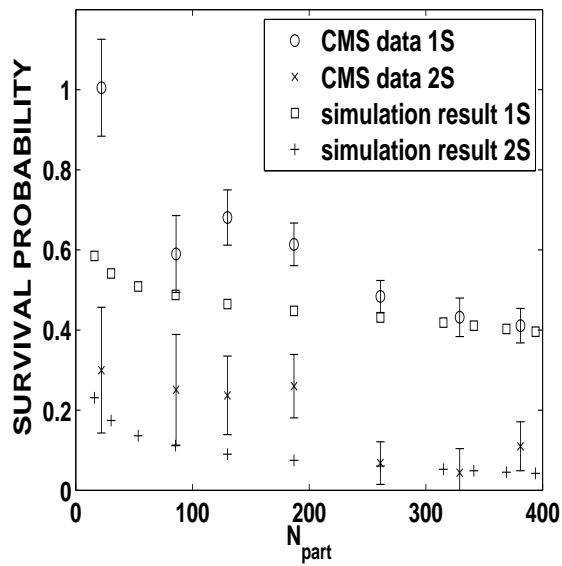}
\caption{CMS data compared with our present simulation results. Both screening and gluonic dissociation along with collisional damping are included. Temperature is calculated according to charge multiplicity.} 
\label{fig:1S_2S_all}
\end{figure}

\begin{figure}[h]
\includegraphics[width = 80mm,height = 80mm]{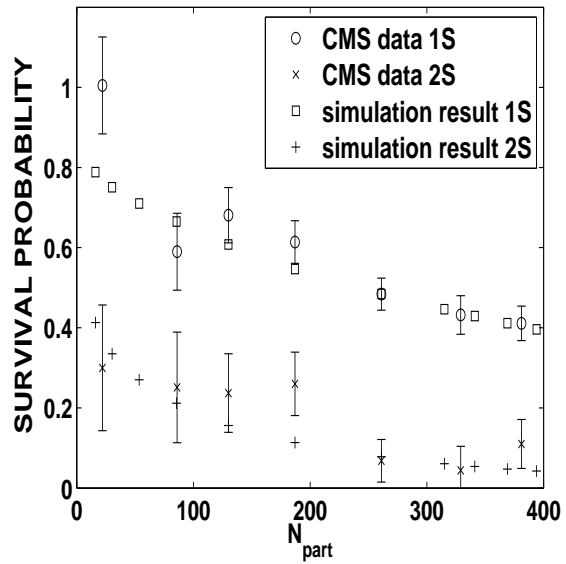}
\caption{CMS data compared with our present simulation results. Both screening and gluonic dissociation with collisional damping are included. The temperature is calculated according to the number of nuclear participants.} 
\label{fig:1S_2S_npart}
\end{figure}

\begin{figure}[h]
\includegraphics[width = 80mm,height = 80mm]{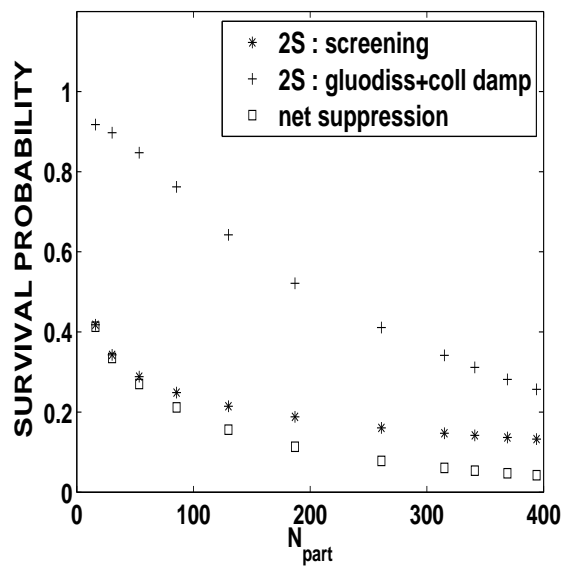}
\caption{Survival probability of $\Upsilon(2S)$ versus $N_{part}$ after including color screening and gluonic dissociation along with collisional damping and including all the above effects based on the $N_{part}$ model of temperature.}
\label{fig:2S_gdiss_npart}
\end{figure}

\begin{figure}[h]
\includegraphics[width = 80mm,height = 80mm]{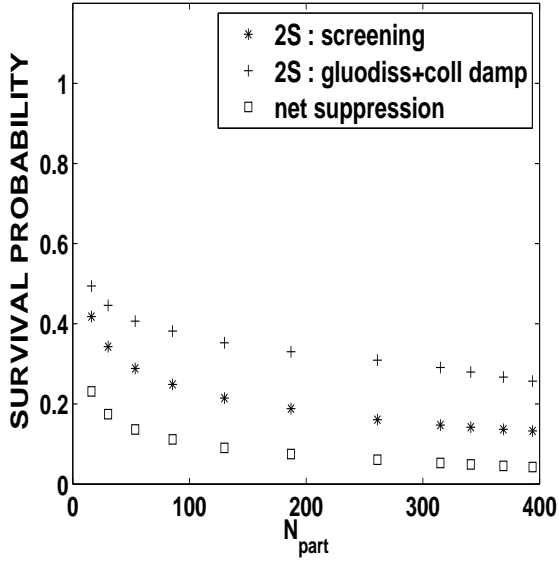}
\caption{Survival probability of $\Upsilon(2S)$ versus $N_{part}$ after incorporating color screening and gluonic dissociation with collisional damping, and including all the above effects based on the multiplicity model of temperature.}
\label{fig:2S_gdiss}
\end{figure}

\begin{figure}[h]
\includegraphics[width = 80mm,height = 80mm]{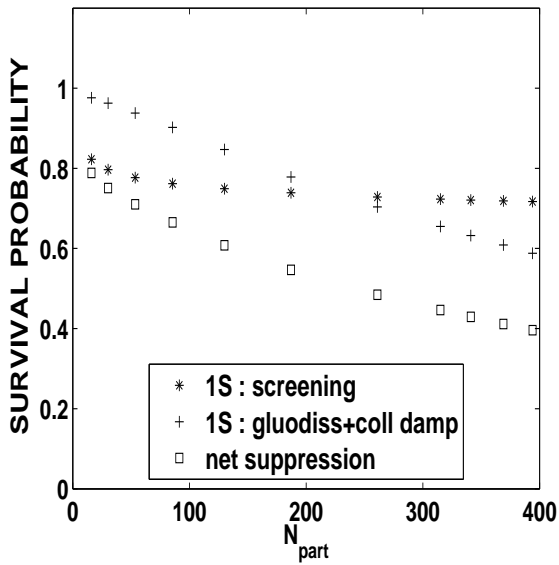}
\caption{Survival probability of $\Upsilon(1S)$ versus $N_{part}$ by incorporating color screening and gluonic dissociation with collisional damping and with all above effects based on the $N_{part}$ variant of temperature model.}
\label{fig:1S_gdiss}
\end{figure}

\begin{figure}[h]
\includegraphics[width = 80mm,height = 80mm]{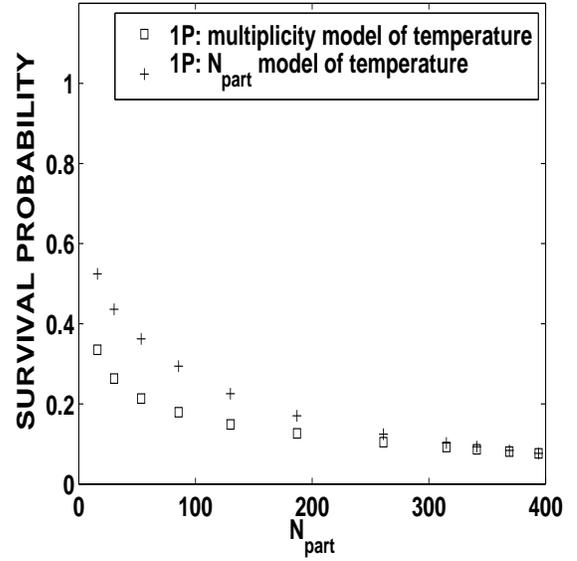}
\caption{Variation of survival probability of the $\Upsilon(1P)$ state with respect to $N_{part}$ including both the temperature models. Color screening and gluonic dissociation along with collisional damping have been incorporated.}
\label{fig:1P}
\end{figure}
\section{Results and Discussions}
{\nd \it Color screening simulation results}:\\
Figure 1 shows the radial wave function of the $\Upsilon$ states, namely $1S$, $2S$, and $1P$. 
Figure 2 depicts the $p_T$ integrated survival probability of $\Upsilon(1S)$ and $\Upsilon(2S)$ states versus number of participants due to color screening alone. We have also shown the CMS suppression data of $\Upsilon(1S)$ and $\Upsilon(2S)$ states for comparison. This figure indicates clearly that screening seems to be able to only partially explain the survival probability of the $1S$ and $2S$ states. In fact, it underestimates the suppression of $1S$ for $N_{part}>150$ but shows better agreement with $2S$ suppression data over the whole centrality region. For $2S$, it predicts slightly lesser suppression in the region $200 \leq N_{part} \leq 350$. The $1S$ and $2S$ simulation results also include feed-down from higher resonance states. The above comparison between our simulated results based on color screening alone and experimental data on $1S$ and $2S$ suppression gives an indication about the role of a new suppression mechanism or more than one mechanism to explain the suppression.
The CMS experimental data as a function of centrality are seen to be nonmonotonic. Although this nonmonotonic behavior is somewhat uncertain and diluted due to the presence of comparatively large error bars in the data, we argue that this trend suggests that at least two different mechanisms play a role. The nonmonotonic region could indicate where one mechanism of suppression ceases to be dominant, while a different mechanism becomes more dominant.\\
{\nd \it Results after combining color screening, gluonic dissociation and collisional damping}:\\
We will now see how the mechanism of gluonic dissociation and collisional damping can provide a potential qualitative explanation for the nonmonotonic behavior of bottomonium suppression. We first consider the case where the temperature used for $S_g$ is modeled using charged particle multiplicity. For this model, the multiplicity values for the four most central bins were linearly extrapolated using ALICE data~\cite{ALICE}. For other bins, the values were directly taken from ALICE data~\cite{ALICE}.
Figure 3 depicts the net suppression for $1S$ and $2S$ after including gluonic dissociation and collisional damping along with color screening.
The temperature used for $S_g$ calculation is modeled by using the multiplicity of charged hadrons measured at chemical freeze-out. The simulation results seem to be overestimating the suppression, particularly for less central collisions ($N_{part}<250$). Also the nonmonotonic region is not very well captured. We now turn our attention to the $1S$ and $2S$ survival probabilities using the temperature model based on $N_{part}$ as shown in Fig. 4. CMS suppression data are also shown for comparison. We see that this temperature model shows better agreement with the CMS data.
We would also like to show that modeling the temperature based on $N_{part}$ results in a better qualitative explanation of nonmonotonicity. Towards this purpose, we show the survival probability versus $N_{part}$ for the $2S$ state in Fig. 5 with color screening and gluonic dissociation along with collisional damping separately. Net suppression has also been plotted on the same plot. We can see that in the temperature model based on $N_{part}$ the gradient of the "+" curve (gluonic dissociation with collisional damping) becomes very different from the "*" curve due to color screening alone in the region $150 \leq N_{part} \leq 200$. This is more or less at the same region where the non-monotonicity occurs in the CMS data for the $2S$ state. 

Figure 6 shows the same plot as Fig. 5 for the $2S$ state but employing temperature modeled by using charged particle multiplicity. The marked change in gradient at around $ 150 \leq N_{part} \leq 200$ (as observed in Fig. 5) is not obtained here. Figure 7 shows the suppression of $1S$ against $N_{part}$ with the color screening scenario alone, with gluonic dissociation and collisional damping alone, and net suppression including all the three effects. Here again temperature is determined by using $N_{part}$. The gradient of suppression for $1S$ due to screening can be seen to flatten out at a lower $N_{part}$ as compared to the $2S$ case. Furthermore, the gradients of the "*" curve (color screening) and the "+" curve (gluonic dissociation plus collisional damping) cross over at around $N_{part}=250$ and they begin to differ significantly in the region $100 \leq N_{part}\leq 150$. 

 In Fig. 8, we finally show the prediction of survival probability of $\Upsilon(1P)$ state with color screening and gluonic dissociation along with collisional damping effects versus $N_{part}$ due to both models of temperature.
With all the above results it seems that temperature based on $N_{part}$ quantitatively and qualitatively depicts better agreement with the CMS data. The temperature model on the basis of multiplicity fails to qualitatively explain the nonmonotonicity in the experimental data. More firm quantitative comparison can be done after normalizing the experimental data by the contribution coming due to cold nuclear matter (CNM) effect,  whose precise value is currently not available for bottomonium at LHC energy. It also needs to be mentioned here that at finite temperature, the spatial extension of the wave functions of the $\Upsilon$ states would be much broader and as a result formation times would be much longer. In the present work, we use a formation time (time taken in the formation of bottomonium bound state, once $Q\bar{Q}$ is produced) which is inversely proportional to the vacuum binding energy of bottomonia~\cite{abdul}. Thus the current model does not take into account the finite temperature effects on the formation times self-consistently during the evolution of the bottomonium after the $Q\bar{Q}$ is formed.

\section{Summary and Conclusions}
We have presented a model of bottomonium suppression in QGP medium by combining the color screening mechanism and gluonic dissociation along with collisional damping. The quasiparticle model is used as an equation of state (EOS) for the quark-gluon plasma (QGP). The model consists of two decoupled mechanisms namely, color screening during bottomonium production followed by gluon-induced dissociation and collisional damping. We further assume that the QGP is expanding according to Bjorken's boost invariant hydrodynamical expansion at mid-rapidity. The final suppression of the bottomonium is calculated as a function of number of participants and the result is compared with recent CMS data at mid rapidity obtained from the CERN LHC. We have found that the current version of the model, based on Debye color screening plus gluonic dissociation along with collisional damping, describes the centrality dependence of bottomonium suppression at LHC energy reasonably well as compared to the color screening model alone. Finally we conclude that the nonmonotonic nature of the CMS data at LHC is an indication that more than one mechanism of suppression come into play in QGP medium. It is worthwhile to mention here that although we have used a phenomenological model to explain the bottomonium suppression at LHC energy and a phenomenology cannot replace explicit theory, the current model shows good agreement with the centrality-dependent LHC data on $\Upsilon$ suppression within experimental uncertainties without varying parameters of the model freely over a large possible range in order to reproduce the data. We have taken values of the parameters that have already been used by earlier researchers.\\ 
{\it Note added in proof.} Recently, we became aware of a related work~\cite{swant}, in which authors have analyzed the data on bottomonium suppression using color screening model alone.
\section{Acknowledgments} 
We thank P. K. Srivastava, et. al., for providing the color screening code. One of the authors (S. Ganesh) acknowledges Broadcom India Research Pvt. Ltd. for allowing the use of its computational resources required for this work. M. Mishra is grateful to the Department of Science and Technology (DST), New Delhi for financial assistance from the Fast Track Young Scientist project.

\end{document}